\documentclass[12pt]{iopart}
\usepackage{graphicx}
\usepackage{dcolumn}
\usepackage{bm}
\usepackage{pstricks}
\usepackage{amssymb}

\newcommand{\ce}{\ensuremath{\varepsilon}}
\newcommand{\bd}{\ensuremath{\bm  {\delta}}}

\begin{document}

\title{Strained graphene: tight-binding and density functional calculations}

\author{R. M. Ribeiro$^1$, Vitor M. Pereira$^2$, N. M. R. Peres$^1$, 
P. R. Briddon$^3$, and A. H. Castro Neto$^2$}

\address{$^1$Department of Physics and Center of Physics, University of Minho,
             Campus de Gualtar, 4710-057 Braga, Portugal,
\ead{ricardo@fisica.uminho.pt, peres@fisica.uminho.pt}}

\address{$^2$Department of Physics, Boston University, 590
Commonwealth Avenue, Boston, MA 02215, USA,
\ead{vpereira@bu.edu, neto@bu.edu}}

\address{$^3$School of Natural Sciences, University of Newcastle upon Tyne,
             Newcastle upon Tyne, NE1 7RU, United Kingdom}

\begin{abstract}
We determine the band structure of graphene under strain using density
functional calculations. The \emph{ab-initio}
band strucure is then used to extract the best fit to the tight-binding hopping
parameters used in
a recent microscopic model of strained graphene. It is found that the
hopping parameters may increase or decrease upon increasing strain, depending on
the orientation of the applied stress. The fitted values are compared
with an available parametrization for the dependence of the orbital
overlap on the distance separating the two carbon atoms. It is also found
that strain does not induce a gap in graphene, at least for deformations up to
10\%.
\end{abstract}
\NJP



\section{Introduction}\label{sec:Intro}

Graphene currently gathers and enormous amount of interest from many fronts.
This stems, mostly, from it being a rare example of a system
whose phenomenology spans a broad specrum of fields. For example, graphene
exhibits many uncommon transport signatures --- as was established during the
earliest experiments following its isolation
\cite{Novoselov:2004,Novoselov:2005,Katsnelson:2006} --- and is an
unexpectedly good conductor, despite being a strictly two-dimensional system.
While graphene has many properties typical of \emph{hard} condensed matter
systems \cite{rmp}, it is simultaneously a soft membrane from
a structural point of view  \cite{Meyer:2007,EunAhKim:2008}. In fact, since
reliable empirical potentials for carbon are generally available, and graphene
isolation is now widely practiced, this system can become a new paradigm in
membrane physics because both accurate microscopic modelling
\cite{Fasolino:2007}, and direct comparison with experiments are possible

The simple fact that graphene is an atomically-thick membrane has a high import
for the interplay between its electronic and mechanical structures. One
particular aspect of this interplay is the extent to which in-plane
strain can modify graphene's electronic structure and, consequently, its
transport characteristics. Strain-induced modifications of the electronic
structure are usually negligible in conventional systems
because of their three dimensional nature. Even in the thinnest films
grown epitaxially on a mismatched substrate, strain is generally irrelevant for
the bulk properties, insofar as it is rapidly and efficiently relieved from
layer to layer above the substrate, either elastically, or by the
intervention of defects \cite{Matthews:1974}. Graphene, on the other hand, is a
single-layer membrane, made out of $sp^2$ hybridized carbon bonds \cite{Euronews}, which are the
strongest in nature. If buckling is disregarded, strain cannot be relived in the
third direction, nor in plane through the generation of defects, which are
energetically costly. This tallies with recent experiments that place graphene
as the strongest material ever measured, when it comes to the elastic response
in the plane of the carbon atoms \cite{Lee:2008}. Graphene can sustain
reversible (elastic) deformations of the order of 20\%, as shown from
\emph{ab-initio} calculations \cite{Liu:2007} and recent
experiments \cite{Kim:2009}. 

Despite these facts, the question of strain and its influence in the electronic
structure of graphene has remained much unexplored until very recently. From
the experimental point of view, important initial steps came from a
series of investigations in the context of Raman spectroscopy \cite{Raman}.
Under in-plane strain the Raman peaks shift considerably, and can be split
under anisotropic deformations. Their dependence with strain can be used to
extract the Gruneissen parameters of graphene, which can be potentially
quite usefull, for one can use a simple Raman measurement to 
\emph{directly} identify and quantify strain profiles in graphene.
Another seminal step was given by Kim and collaborators, who have
investigated transport properties of graphene under strain, achieved by
depositing samples on stretchable substrates \cite{Kim:2009}.

From the theoretical front, a vital question is whether small and
easily achievable strain can induce a bulk spectral gap in graphene's spectrum.
If so, this would have enormous consequences in the context of a graphene
device with tunable electronic structure. An early density functional
theory calculation (DFT) \cite{Gui:2008} advanced that any arbitrarily small
amount of uniaxial strain opens a gap in graphene's spectrum, whose magnitude
varies non-monotonically with the amount of strain. These findings were
apparently seconded by another \emph{ab-initio} calculation \cite{Ni1},
although there was strong disagreement between the magnitude of the gap among
those two calculations, for the same amount of strain.
Recently, however, Pereira and collaborators \cite{PeresStrain}, using a
tight-binding approach and treating deformations within linear continuum
elasticity, challenged those conclusions. They
showed that a spectral gap is achievable only for uniaxial deformations in
excess of 20\%, and that the effect strongly depends on the direction of the
deformation with respect to the underlying lattice. These results are
consistent with the investigations of Hasegawa \etal\ which
show that the gappless Dirac spectrum is robust with respect to arbitrary
(and not exceedingly large) changes in the nearest-neighbor hoppings
\cite{Hasegawa:2006}. Recent developments from the \emph{ab-initio} front
\cite{Ni2,farjam2009} have shown results in agreement with the gappless scenario
put forth in reference \cite{PeresStrain}. The apparent contradiction among 
different \emph{ab-initio} calculations can be traced to the peculiarities of
the electronic spectrum in graphene. In particular, the fact that, under strain,
the Dirac point drifts away from the high-symmetry points of the lattice 
\cite{PeresStrain} was overlooked in the interpretation of the earliest
results, and led to the erroneous conclusion that a gap seemed possible for any
amount of strain.

Given that strain is now perceived as a new avenue of research in the physics of
graphene, and given the importance of simple microscopic models that reliably
describe the evolution of the electronic system under strain, we intend to
further clarify these issues by pursuing two complimentary goals. We perform an
\emph{ab-initio} calculation of the band structure of graphene under uniaxial
strain, for deformations up to 10\%. The calculated bandstructure allows us to
establish the absence of a spectral gap in the spectrum. Subsequently, we fit
the tight-binding model used in reference \cite{PeresStrain} to the
bandstructure obtained here \emph{ab-initio}, in order to ascertain its range of
validity, and to extract the model parameters. We conclude that the
parameterization for the hopping integrals used in the cited reference is
generally applicable in the entire range of deformations used in our study.

This paper is organized as follows. We start the next section by discussing the
general features of strain in the honeycomb lattice, and the tight-binding
parameterization that will be fit to our \emph{ab-initio} bandstructure. In
Sec.~\ref{sec:DFT} we present the details of our DFT calculations, followed,
in Sec.~\ref{sec:Bandstructure}, by the procedure used here to study the
electronic structure as a function of strain. The calculated banstructures
and their fitting to the tight-binding dispersion are shown and discussed in
Sec.~\ref{sec:TBFit}. In Sec.~\ref{sec:TBparameters} we analyse the variation
in the tight-binding hopping integrals, as fitted to the \emph{ab-initio}
bands, and compare their strain dependence with the analytical form proposed
in reference \cite{PeresStrain}.

\section{General considerations on deformed graphene}\label{sec:Genesral}

\begin{figure}
  \centering
  \includegraphics*[width=0.5\columnwidth]{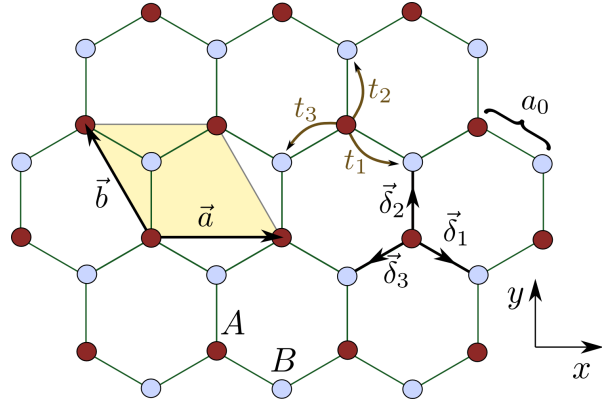}
  \caption{(Color online) 
    Illustration of the honeycomb lattice with the $A$ and
    $B$ sublattices, the lattice vectors $\bm\delta_i $ ($i=1,2,3$), and the
    hopping parameters $t_1$, $t_2$ and $t_3$.
    The abscissas are along the zig-zag edge (horizontally in the figure).
    Also shown are the primitive vectors $\bm a$ and $\bm b$ used in the DFT
    calculation, and $a_0$ is the equilibrium carbon-carbon distance.
  }
\label{fig:unitcell}
\end{figure}

In Fig.~\ref{fig:unitcell} we represent the unit cell of graphene, depicting the
next nearest-neighbor vectors, $\bm\delta_i$ ($i=1,2,3$), and
the hopping parameters, $t_i$. The primitive vectors, $\bm a$ and $\bm b$,
used in the DFT calculations are also shown.
In this study we consider only two types of uniaxial strain:
(i) along the $x$ direction -- corresponding to strain parallel to
the zig-zag edge of the ribbon; 
(ii) along the $y$ direction --
corresponding to strain along the armchair edges of the ribbon. 
These correspond to two particular orientations of the more general uniaxial
case discussed in Ref.~\cite{PeresStrain}, where an arbitrary orientation
with respect to the lattice was considered. For small strain (appropriate for
our study), the length of the vectors $\bm\delta_i$  (in units of $a_0$)
is given by \cite{PeresStrain}
\begin{eqnarray}
\label{eq:DeformedBonds-ZA}
  |\bd_1| &=&|\bd_3| = 1 + \frac{3}{4}\ce -\frac{1}{4}\ce\sigma
  \,,\;
  |\bd_2| =1 - \ce\sigma\nonumber
  \\
  |\bd_1| &=&|\bd_3| = 1 + \frac{1}{4}\ce -\frac{3}{4}\ce\sigma
  \,,\;
  |\bd_2| =1 + \ce\,,
\end{eqnarray}
for zig-zag and armchair deformations, respectively. 
In our notation, $\ce $ represents the longitudinal strain and $\sigma=0.165$ is
the Poisson ratio, as measured for graphite \cite{black}, or $\sigma=0.10-0.14$
for graphene as calculated in Ref. \cite{farjam2009}. 
It is clear from Eq. (\ref{eq:DeformedBonds-ZA})
that both $t_1$ and $t_3$ will change upon stress by the same value, since the
corresponding change in $\delta_1$ and $\delta_3$ is the same. In Ref.
\cite{PeresStrain} it was found that: 
\begin{enumerate}
  \item For stress along the zig-zag edge, $t_1$ and $t_3$ decrease and
        $t_2$ increase upon increasing strain.
  \item For stress along the armchair edge, all $t_i$ decrease, with
        $t_2$ being smaller than $t_1$ and $t_3$.
\end{enumerate}
These findings result from a combination of Eq.~(\ref{eq:DeformedBonds-ZA})
with a parametrization for the change of the hopping parameters with the bond
length given by \cite{papa}
\begin{equation}
  V_{pp\pi}=t_0e^{-\beta_i(l/a_0-1)}\,,
  \label{vpppi}
\end{equation}
where $t_0$ is the hopping integral in free-standing graphene, $l$ the
bond length, and $\beta_i$ a number of order $\beta_i\sim 3$. 
One of our goals is to verify to which extent this parametrization
(\ref{vpppi}) is valid, starting from a full DFT calculation of graphene's bands
under stress. With that purpose, we shall compare quantitatively the
above parametrization for the variation of $t_i$ with the values of
$t_i$ obtained from adjusting the tight-binding and DFT bands. 
For our DFT calculations it is convenient to write the
primitive vectors of the unit cell as (see Fig. \ref{fig:unitcell}):
\begin{eqnarray}
  \bm{a} &=& a\;\bm{e_x}\,,\\
  \bm{b} &=& -\frac{a}{2}\;\bm{e_x} + \frac{\sqrt{3}}{2}b\;\bm{e_y}\,, 
\end{eqnarray}
The parameter $a$ was varied when stress was applied along the $x$ axis,
and likewise for $b$, when stress is applied along the $y$ direction.

\section{Details of the DFT calculations}\label{sec:DFT}

In our study of the spectrum of graphene under stress, density functional (DFT)
calculations were performed with an {\it ab-initio} spin-density functional code
({\sc aimpro})\cite{Rayson2008}, along with the local density approximation
(LDA).

The Brillouin-zone (BZ) was sampled for integrations according to the scheme
proposed by Monkhorst-Pack\cite{Monkhorst1968}. 
A grid of $12\times12\times1$~$\bm{k}$-points was generated and folded according
to the symmetry of the BZ. An increase in the number of points did not result in
a significant total energy change.
However, a careful choice of the sampled $\bm{k}$-points is necessary in this
study (see more below).

We use pseudo-potentials to describe the ion cores.
Lower states (core states) are accounted for by using the dual-space separable
pseudo-potentials by Hartwigsen, Goedecker, and Hutter\cite{Hartwigsen1998}. 
The valence states are expanded over a set of $s$-, $p$-, and $d$-like
Cartesian-Gaussian Bloch atom-centered functions, and the states are filled
according to the  Fermi-Dirac distribution using a value of $k_B T =0.01$
{\ttfamily eV}, a procedure known to accelerate the convergence of the
calculations. 
Kohn-Sham states are expressed as linear combinations of these basis functions,
which were optimized for graphite.

Graphene was modeled in a slab geometry by including a vacuum region in a unit
cell containing 2 carbon atoms. 
In the normal direction ($z$-direction), the vacuum separating repeating slabs
has more than 30 \AA ($c=31.75$ \AA). 
The size of the cell in the $z$-direction was optimized to make sure there was
no interaction between repeating slabs. 
The size of the unit cell in the plane direction was optimized, and the lattice
parameter after relaxation is $a=2.4426$ \AA. The tolerance for stopping
structural optimization was $10^{-6}$~Ha. The tolerance for self-consistency was
$10^{-6}$~Ha.

\section{Bandstructure calculations under strain}\label{sec:Bandstructure}

Our calculations implement the deformation of the lattice along the following
steps: 
the unit cell of graphene was first strained in the $x$ direction and
no relaxation was first allowed in the $y$ direction, which is at first sight a
reasonable approximation if the strain is small (this hypothesis is confirmed
by the DFT calculations).
This is validated by the small Poisson ratio for graphene, $\sigma\sim
0.10-0.14$, calculated in Ref.~\cite{farjam2009} for much larger strains. Latter,
we have  allowed the lattice to relax along the 
$y$ axis, probing the energy
landscape as a function of different bond lengths along $y$ thus locating in this way
the energy minimum of the relaxed lattice. 
The reasons for studying both the relaxed and unrelaxed lattice are given below. 
This allowed us to
extract a Poisson ratio of $\sigma\sim  0.13-0.15$, in agreement with the calculations of
Ref. \cite{farjam2009}.
The band structure of strained graphene was then calculated using DFT, for
a fixed value of $\epsilon$. The resulting DFT valence band around the ${\bm
K}-$point was subsequently used to find the best values of $t_i$ that fit the
tight-binding bandstructure. Our choice of the valence band to fit the hopping
parameters is motivated by the documented lack of accuracy of DFT in describing
empty states.
In the fitting procedure the bands were cut-off at 0.2 eV, well inside the
validity of the Dirac cone approximation for unstrained graphene. 
A fit for the values of $t_i$ valid over the full energy range of the DFT
graphene bands was found to be unsatisfactory using the 
simple Eq. (\ref{eq:TB}). This is not surprising because Eq. (\ref{eq:TB})
neglects details like the overlap factors of the orbitals, and other details
discussed in Ref.~\cite{pablo}.
In addition, the expression used for the tight-binding energy includes only
hopping to the first neighbors, albeit with different values for the parameters
$t_1$, $t_2$ and $t_3$\cite{rmp}:
\begin{equation}
  E = \pm \vert t_2 + t_3 e^{-i \bm{k}\cdotp\left( \bm{a}+\bm{b}\right) } + 
    t_1 e^{-i \bm{k}\cdotp\bm{b}}\vert\,.
  \label{eq:TB}
\end{equation}
If the lattice is strained only along the two chosen directions -- $x$ and
$y$ --, symmetry imposes that $t_3=t_1$. 
The above procedure was then repeated for stress along the armchair direction.

\begin{figure}
  \centering
  \includegraphics*[width=0.7\columnwidth]{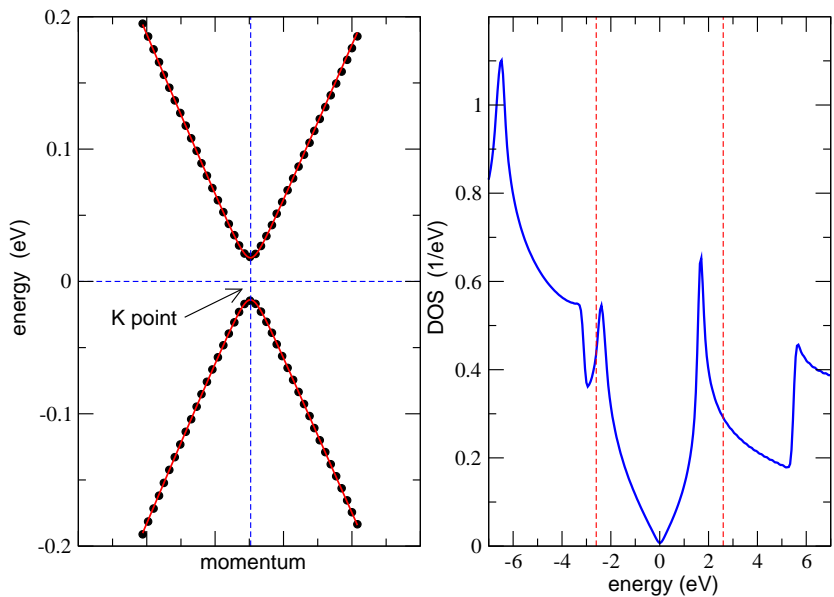}
  \caption{(Color online) 
    Fitting of the DFT data (circles) to the tight-binding expression in Eq.
    (\ref{eq:TB}) (solid line), for 1\% strain (deformation along the
    $x$-direction). The fit is performed around the $\bm{K}$ point in the BZ,
    and along the $\Gamma-\bm{K}$ direction. 
    Note, however, that under strain the bands do not touch at the $\bm{K}$
    point anymore \cite{PeresStrain}. The system remains gapless, as can
    be seen on the right panel, where the density of states (DOS),
    as computed from DFT data for the same strain as in the left panel, is
    given.
    Also depicted in the right panel are the positions of the two Van Hove
    singularities in unstrained graphene (dashed lines).
  }
  \label{fig:fit}
\end{figure}

\section{DFT versus tight-binding}\label{sec:TBFit}

Figure \ref{fig:fit} presents a fitting of Eq. (\ref{eq:TB}) to the DFT results
(points), together with the fitting (solid line)
of the tight-binding Eq.  (\ref{eq:TB}). 
The fit of the numerical data to Eq. (\ref{eq:TB}) was done for momenta around
the ${\bm K}$ point for all the values of strain. As can be seen in Fig.
\ref{fig:fit}, for finite strain the touching of the valence and conducting
bands does not happen at the ${\bm K}$ point. This was shown explicitly in Ref.
\cite{PeresStrain} and, as a result, any plot of the bandstructure in the close
vicinity of $\bm{K}$ will inevitably show a fictitious gap. In reality the
system remains gapless, as can be seen from the DFT density of states plotted in
Fig. \ref{fig:fit}.

To fit the tight-binding dispersion we used twenty LDA points from
each side of the ${\bm K}$ point in the direction $\Gamma$--${\bm  K}$. As 
mentioned earlier, the fit was done only for the valence band,
although Fig.~\ref{fig:fit} also shows the DFT data for the conduction band
together with the tight-binding spectrum using the values of $t_i$ from the
fit to the valence band. The agreement is excellent.
The 41 momentum points used for the fitting span a reciprocal length in momentum
space of the order of $0.08$~rad/bohr.
A few notes are worth to be cast here. As strain is induced in graphene, the
hexagonal symmetry is lost and the ${\bm  K}$ points do not retain their
original position in the Brillouin zone. As an example, one of those symmetry
points lies at the position given by the general expression:
\begin{eqnarray}
  {\bm K}&=& 
  \left( \frac{c_{1y}}{2}\frac{c^2_{2x} + c^2_{2y} - c_{1x}c_{2x} - 
c_{1y}c_{2y}  }{c_{1y}c_{2x} - c_{1x}c_{2y}  }+\frac{c_{1x}}{2}\right. ,
\nonumber\\
&-&
\left. \frac{c_{1x}}{2}\frac{c^2_{2x} + c^2_{2y} - c_{1x}c_{2x} - 
c_{1y}c_{2y}  }{c_{1y}c_{2x} - c_{1x}c_{2y}  }+\frac{c_{1y}}{2} \right)
\,,
\end{eqnarray}
where
\begin{eqnarray}
 {\bm c_1} = \left( c_{1x}, c_{1y} \right)\,,\\
{\bm c_2} = \left( c_{2x}, c_{2y} \right)
\end{eqnarray}
are the primitive vectors of the Brillouin zone, associated with the
distorted unit cell (see Fig. \ref{fig:unitcell}).
We have calculated the coordinates of the ${\bm  K}$ point for each value of the
strain, and verified that the valence and conduction bands do not touch each
other at this point. As found previously in \cite{PeresStrain} using a
tight-binding approach, the ${\bm  K}$ point and the point in momentum space
where the valence and conduction bands touch do not coincide. Our calculations
show no gap opening in graphene, which agrees with the calculations of
the cited reference, and also the DFT calculations of Ref. \cite{farjam2009}.

This point is indeed crucial for the discussion of the bandstructure under
strain, since computing the spectrum around the ${\bm  K}$ point only may lead
to the erroneous conclusion that strain opens a spectral gap \cite{Ni1}, a
fact not supported by a more detailed analysis \cite{Ni2,farjam2009}, and our
current results. DFT methods, inevitably use a finite grid of momentum
values over the Brillouin zone, which are used to sample the spectrum
and the corresponding density of states. Using a too coarse sampling of the
Brillouin zone is most likely bound to miss the precise momenta at which the
valence and conduction bands touch. This would produce a density of states
featuring an artificial non-existing gap, a consequence of an aliasing effect
\cite{Ni1}.

\begin{figure}
  \centering
  \includegraphics*[width=0.7\columnwidth]{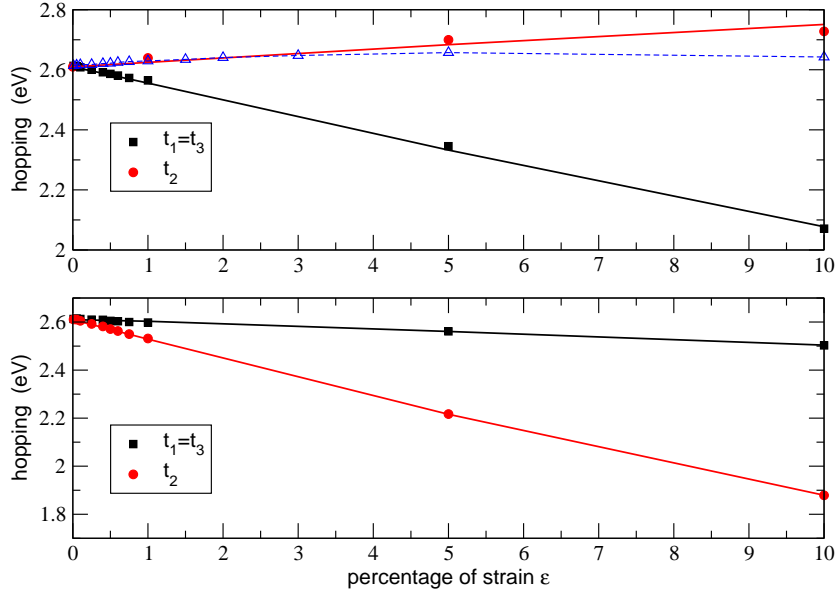}
  \caption{(Color online) 
    Variation of hopping parameters, $t_1=t_3$ and $t_2$, 
    as function of the strain $\ce$  determined from fitting Eq. (\ref{eq:TB})
    to the DFT valence band (points). 
The upper panel shows the case where the
    lattice is deformed along the zig-zag direction, while the lower panel 
    refers to strain along the armchair direction. The solid lines represent
    the hopping computed using Eqs. (\ref{eq:DeformedBonds-ZA})
    and (\ref{vpppi}) and the parametrization given in
Table \ref{tab-beta}. The dashed line with triangles in the upper panel
is the value of $t_2$
when the length of the corresponding bond is kept constant.
    The maximum amount of strain was 10\%.
 The error bars are of the size of
    the points.
  }
  \label{fig:txty}
\end{figure}

\section{Tight-binding hopping parameterization}\label{sec:TBparameters}

Figure \ref{fig:txty} shows the variation of the hopping parameters as graphene
endures stress along the zig-zag direction (upper panel).  
We have strained graphene's unit cell up to 10\%, although experiments seem
to indicate that the material can support reversible strains up to 20\%
\cite{Liu2007}.
The hopping parameter $t_2$ is perpendicular to this direction, and according to
Eqs.
(\ref{eq:DeformedBonds-ZA}) should have, in this case, a small variation due to the
small value of the Poisson coefficient.
Consequently, we have studied two cases for stress along the 
zig-zag direction: (i) keeping constant the bond distance associated
with $t_2$; (ii) allowing this distance to vary, such that the energy of the
strained lattice is minimum. The study (i) allow us to discuss whether the change in 
$t_2$ is only due to the bonding length modification or is also controlled by the
redistribution of the electronic density around the carbon atoms.
 It is worth
noticing that, according to the simple tight binding description of Eqs.
(\ref{eq:DeformedBonds-ZA}) and (\ref{vpppi}), the
change in the value of $t_2$ is due to the modification of the bond length
alone, a result not observed in our DFT calculations, where $t_2$ varies with
strain, even under the approximation of keeping the corresponding bond length
constant (the variation of $t_2$ is, nevertheless, very small).
Additionally, as  graphene is strained along the zig-zag direction the hopping
parameters $t_1$ and $t_3$ decrease, certainly due to the change of the bond
length associated with these parameters.
The overall results can be understood as follows: strain along the zig-zag
direction increases the bond length along $t_1$ and $t_3$ directions and
reduces the electronic density along these same bonds,
additionally  it increases the electronic density on the bond length
associated with $t_2$, even if no deformation is allowed for this bond.
Consequently, $t_1$ and $t_3$ are reduced and $t_2$ increases slightly. This
redistribution of electronic density among the several bonds is effectively
included in the tight binding description by allowing  a change of all bond
lengths, but this is not strictly necessary to observe the effect. We are then
forced to conclude that the change in $t_2$ stems from a combination of the two
effects: electronic density redistribution among the bonds and change in the
bond length.
 This is in line with the fact that the relative orientation of
the orbitals is also affected by the deformation and the resulting
re-hybridization alone contributes to a modification of the effective
hopping, even if the bond length remains unmodified. It is worth noticing that
the values of $t_2$ for the relaxed and unrelaxed lattice are essentially the
same for strain up 3\%, as seen in the upper panel of Fig. \ref{fig:fit}.
The points for strains below 1\% were obtained without relaxation in the 
direction perpendicular to the strain.
The agreement of these points to the parametrization confirms our 
assumption that the relaxation is not important for small values of strain.

The situation is different for strain applied along the armchair direction
(see lower panel of Fig. \ref{fig:txty}), because in this case all three
bonds are deformed, up to first order in $\ce$ without any contribution from the
Poisson coefficient, as can be seen from Eq. (\ref{eq:DeformedBonds-ZA}).
Since the bond associated with $t_2$ decreases considerably more than the other
two, this hopping decreases faster upon strain, an effect seen in Fig.
\ref{fig:txty}.

In either the zig-zag or armchair cases, the bandstructure in the neighborhood
of the neutrality point is seen to be well described by the
parametrization used in the tight-binding analysis of Ref. \cite{PeresStrain},
and given by Eq. (\ref{vpppi}). This fact is documented by the agreement
between the solid lines and points in both panels of Fig.~\ref{fig:txty}
In Table \ref{tab:hopping} we present the values of the parameter $\beta_i$ 
[Eq. (\ref{vpppi})] associated with each bond, extracted for the different cases
studied here.

\begin{table}
  \label{tab-beta}
  \centering
  \begin{tabular}{c|c|c}
    \hline
    Stress & $t_i$ & $\beta_i$ \\
    \hline
    $x$-direction & $t_1=t_3$ & 3.15 \\
        & $t_2$ &  4.0\\
    \hline
    $y$-direction & $t_1=t_3$ & 2.6 \\
        & $t_2$ & 3.3\\
    \hline
  \end{tabular}
  \caption{
    Results for the parameter $\beta_i$ in Eq. (\ref{vpppi}) for
 stress along the zig-zag ($x-$direction) and armchair ($y-$direction)
directions, both considered in the text.}
  \label{tab:hopping}
\end{table}


\section{Conclusions}\label{sec:Conclusions}

We have calculated the bandstructure of graphene under uniaxial strain
\emph{ab-initio}, within the LDA approximation. The spectrum remains gapless
for all strain configurations studied, and up to the maximum value of
longitudinal deformation (10\%) used in our calculations, tallying with recent
similar investigations \cite{Ni2,farjam2009}.
The \emph{ab-initio} bandstructures were used to fit a tight-binding
parametrization of the dispersion, from where we
extracted the effective nearest-neighbor hopping parameters, and their
dependence with the magnitude of deformation.
As is generally known, hopping parameters calculated using DFT are
lower than the experimental ones, which is also seen in our
calculations that show an unstrained hopping of $2.6\,$eV.
Nevertheless DFT is accurate when calculating energy differences, and
thus we believe that the slopes $\beta_i$ of the hopping parameters calculated
above should be close to the experimental values.
Moreover, the behavior seen for the dependence of $t_i$ on $\ce$ follows
rather well the trend given by Eq.~(\ref{vpppi}). We expect that the
results found for $\beta_i$ can be extrapolated to large values of $\ce$ with a
certain degree of confidence.

\ack
We wish to acknowledge the support of the Funda\c{c}\~ao para a Ci\^encia e a
Tecnologia (FCT) under the SeARCH (Services and Advanced Research Computing with
HTC/HPC clusters) project, funded by FCT under contract CONC-REEQ/443/2005.
VMP and NMRP are supported by FCT via grant reference PTDC/FIS/64404/2006. 
AHCN acknowledges the partial support of the U.S. Department of Energy
under the grant DE-FG02-08ER46512.


\end{document}